\documentclass[11pt,a4paper,graphicx ]{article}
\usepackage{slashed} 
\usepackage{graphicx}
\usepackage{epsfig}
\usepackage{multirow} 
\usepackage{epstopdf}
\usepackage{setspace}
\usepackage{amsmath,amssymb,mathtools}
\usepackage{subfig}

\usepackage{jheppub}  
\usepackage{graphicx,color}
\usepackage{amsmath}
\usepackage{mathtools}
\usepackage{varwidth}
\usepackage{amsfonts,fontenc}
\usepackage{hyperref}
\usepackage{ragged2e}
\usepackage{xr}

\newcommand{\alspi}{\ \frac{2\alpha_s}{3\pi}}

\newcommand{\ba}{\begin{align}}
\newcommand{\ea}{\end{align}}

\title{ Next-to-leading power corrections to spherocity distribution at NLO in QCD} 

\author[]{Shubham Mishra}

\affiliation[]{Department of Physics, Indian Institute of Technology Hyderabad, Kandi, Sangareddy, Telangana State 502285, India}

\emailAdd{shubhamhere82@gmail.com}

\abstract{This study explores the leading contributions at the next-to-leading order for the event shape variable, spherocity. Our investigation is presented through a combination of analytical derivations and graphical representations. Additionally, we delve into the intriguing behavior of the spherocity distribution, mainly as it arises from different regions of the allowed phase space. }

\begin{document} 
\maketitle
\preprint{}
\newcommand{\setfmfoptions}[0]{\fmfset{curly_len}{2mm}\fmfset{wiggly_len}{3mm}\fmfset{arrow_len}{3mm} \fmfset{dash_len}{2mm    }\fmfpen{thin}}
\def\ft#1#2{{\textstyle{\frac{#1}{#2}}}}
\pagebreak

\section{Introduction}
Event shape variables are observables that are capable of studying the geometrical features of energy flow in QCD events. They were among the first observables to be used to test the QCD predictions.
These studies have been instrumental in determining the value of the strong coupling constant ($\alpha_{s}$)~\cite{Gehrmann-DeRidder:2007nzq, Becher:2008cf, Dasgupta:2003iq,
	Catani:1991kz}, studying power corrections~\cite{Caola:2021kzt, Caola:2022vea, Nason:2023asn, Gardi:2001di, Agarwal:2020uxi}, extraction of quark and gluon color factors, and beta function.
Moreover, examining event shape variables confirmed that gluons are indeed vector particles, this conclusion contradicted earlier theoretical predictions that treated gluons as scalar particles, which did not align with experimental observations~\cite{Ellis:1991qj}. Event shape observables have proved to be an essential tool to study the power corrections~\cite{Beneke:1998ui} and hadronization effects. 
A ubiquitous and exciting feature of shape variables is related to the appearance of large logarithmic corrections from the regions where the event shape value is small, as these are primarily the regions where we have soft gluon emissions. These emissions have larger emission probability in comparison to hard gluon emissions. 
The theoretical uncertainty arising in  $ \alpha_s $ determination from event shape studies~\cite{ParticleDataGroup:2018ovx, ParticleDataGroup:2020ssz} are much higher than the low energy computations from lattice QCD~\cite{FlavourLatticeAveragingGroupFLAG:2021npn, dEnterria:2022hzv}, which requires the need of further theoretical progress in the area of event shape studies.  
To address this issue, new ideas are being developed; one such important direction of progress is the study of soft drop techniques~\cite{Larkoski:2014wba, ATLAS:2017zda, Marzani:2017kqd, Kang:2018vgn, Hannesdottir:2022rsl} and grooming of jet observables.

The shape variables thrust~\cite{Farhi:1977sg}, $ C$-parameter~\cite{Donoghue:1979vi, Fox:1978vu, Parisi:1974sq, Ellis:1980wv}, jet broadening~\cite{Rakow:1981qn, Ellis:1986ig, Catani:1992jc} and jet masses~\cite{Salam:2010nqg, Dasgupta:2013ihk} have received much more attention than spherocity~\cite{Georgi:1977sf} and sphericity~\cite{Ellis:1976uc}, although these two observables were defined around the same time as others. Sphericity turned out to be infrared unsafe, and theoretical predictions involving infrared-unsafe observables are known to be plagued with infrared singularities. On the other hand, spherocity is an infrared-safe observable; thus, the prediction involving spherocity will be free of infrared singularities and trustworthy. However,
spherocity and transverse spherocity has recently found many popular applications in study of heavy-ion collisions~\cite{Ortiz:2023slm,  Cuautle:2014yda, BelloMartinez:2019fnp, Prasad:2022zbr, Rath:2024ljf, Khatun:2019dml, Mallick:2020dzv, Ortiz:2020rwg, Deb:2020ezw, Mallick:2020ium, Mallick:2021wop, Mallick:2021yky, Oliva:2022rsv}.
Studying the spherocity distribution in QCD is essential for improving our theoretical understanding of QCD dynamics and making precise predictions that can be compared with experimental data. These calculations provide valuable insights into the behavior of QCD at high energies and help test the theory's validity in different kinematic regimes. Initial studies performed with spherocity to study the jet structure can be found in~\cite{Herquet:1980gd, Goggi:1979uq, Binetruy:1979ie, Bopp:1979yp, Gottschalk:1979dq, Rizzo:1979mx, Saclay:1979fer, DeRujula:1978vmq}.
This work aims to derive and analyze the spherocity distribution at next-to-leading order in $ \alpha_{s} $.

To be more specific, if $\xi$ is a dimensionless kinematic variable, such
that $\xi\rightarrow 0$ towards the elastic region, the corresponding differential cross-section has the generic form

\begin{equation}
	\frac{d \sigma}{d \xi} \, \sim \, \sum_{n = 0}^{\infty} \left( \frac{\alpha_s}{\pi} \right)^n \, 
	\left[   \sum_{m = 0}^{2 n - 1} \left\{ c_{n m}^{\text{LP}}
	\left( \frac{\log^m \xi}{\xi} \right)_+   + \, c_{nm}^{\text{NLP}} \, \log^m \xi \, + \, \ldots \right\} +
	c_n^{(\delta)} \delta(\xi) \right] \, ,
	\label{eq:distribution}
\end{equation}
without any loss of generality we can take threshold limit as  $ \xi \to 0 $.
The first term on the right in above expression is known as the leading power (LP) term, under the threshold limit LP term is the most significantly contributing term in the entire distribution. Further the LP term is made of various powers of logarithms known as leading log (LL), next-to-leading log (NLL), next-to-next-to-leading logs (NNLL) so on and so forth, within the LP the LL is the most significant term. Much has been known about the LP term to arbitrary order and there have been numerous approaches towards their resummation~\cite{Parisi:1979xd, Curci:1979am, STERMAN1987310, Catani:1989ne,
	Catani:1990rp, Gatheral:1983cz, Frenkel:1984pz, Sterman:1981jc,
	Korchemsky:1992xv, Korchemsky:1993uz,Forte:2002ni,
	Contopanagos:1996nh, Becher:2006nr, Schwartz:2007ib,
	PhysRevD.78.034027, PhysRevD.80.094013} and  power correction studies~\cite{Gardi:2001di, Agarwal:2020uxi}. Reviews on some of
these different approaches can be found in~\cite{Laenen:2004pm,Luisoni_2015,Becher:2014oda,Campbell:2017hsr,Agarwal:2021ais}. The LL at LP is very well known to originate from soft and collinear radiations while the subsequent suppressed logarithm the NLL receives contribution from collinear emissions. The second term on right is set of terms proportional to $ \xi^0 $ and known as next-to-leading power (NLP) terms, although they are suppressed by a power of $ \xi $, are known to be relevant, as they grow logarithmically towards threshold. These terms originate from various sources, such as next-to-soft gluon emission, soft fermionic emission and others. In contrast to the LP terms, the precise organization of these
NLP terms to all orders and arbitrary logarithmic accuracy are not yet clear. Patterns among NLP terms have been studied for various processes~\cite{Bonocore2015, Bonocore2016, Moult:2018jjd, Beneke:2019oqx,
	Moult:2019mog, Bahjat-Abbas:2019fqa, Beneke:2019mua, Moult:2019vou,
	Ajjath:2020sjk, Beneke:2020ibj, Ajjath:2020ulr, vanBeekveld:2021mxn,
	Liu:2019oav,Pal:2023vec,vanBeekveld:2023gio,Ferrera:2023vsw,Balsach:2023ema}. For a number of observables
their numerical contribution can be significant~\cite{KRAMER1998523, BALL2013746,PhysRevLett.114.212001,
	vanBeekveld:2019prq, vanBeekveld:2021hhv, Ajjath2022}.
The all-order resummation of NLP terms has been pursued through
different approaches, such as a diagrammatic and a path integral approach given
in~\cite{Laenen_2009, Laenen2011}, while a physical kernel
approach is pursued in~\cite{Soar2010OnHD, Florian2014ApproximateNH,
	Presti2014LeadingLL}.
From direct QCD formalism, the development of factorization
theorems for NLP terms extending from LP terms are studied in~\cite{Bonocore2015,Bonocore2016,Bonocore2020AsymptoticDO,PhysRevD.95.125009,PhysRevD.96.065007,Gervais2017SoftRT,PhysRevD.103.034022, Vernazza:2023hrf, Engel:2023rxp} (for NLP factorization see,~\cite{Beneke:2022obx}).
The study of NLP effects in the framework of SCET has many aspects,
involving also operators contributing at NLP level~\cite{Kolodrubetz:2016uim, Moult:2016fqy, Feige:2017zci, Beneke:2017ztn, Beneke:2018rbh, Bhattacharya:2018vph, Beneke:2019kgv, Bodwin:2021epw}, the development of factorization~\cite{Moult:2019mog, Beneke:2019oqx, Liu:2019oav, Liu:2020tzd}, and the explicit studies of physical observables~\cite{Boughezal:2016zws, Moult:2017rpl, Chang:2017atu, Moult:2018jjd, Beneke:2018gvs, Ebert:2018gsn, Beneke:2019mua, Moult:2019uhz, Liu:2020ydl, Liu:2020eqe, Wang:2019mym, Beneke:2020ibj}.  Reviews on some
these different approaches can be found in~\cite{Laenen:2004pm,Luisoni_2015,Becher:2014oda,Campbell:2017hsr,Agarwal:2021ais}, new QCD ideas~\cite{Banfi:2001bz,Catani:1992jc,Catani:1992ua,Ellis:1991qj},
e.g. the development of resummation~\cite{Catani:1992ua} and fixed-order studies.

This paper will analytically compute the spherocity distribution at one loop for the process $ e^+ e^- \to q \, \bar{q} \, g $. Our primary focus is on the LLs and NLLs that appear in LP and NLP terms, aiming to understand their origin. This paper is structured as follows: in section~\ref{sec:spherocity}, we re-introduce the shape variable spherocity and the details about the process we are concerned with; in section~\ref{sphero_NLO}, we compute spherocity distribution at NLO by breaking the contributions into three different regions of phase space. In section~\ref{sec:PHASEspacePLOTS}, we analyze and discuss the unique features of spherocity distributions through analysis of constrained phase space plots, we conclude our findings in section~\ref{sec:conclusion}.

\section{Spherocity}\label{sec:spherocity}

Spherocity is an event shape variable defined to study the jet structure and the geometry of final state events, and it is an infrared-safe shape variable, i.e., it is insensitive to soft emissions or
collinear splittings. For $ n $ particle final state, it is defined as 

\begin{align}
	s_p= \text{min}\left(\frac{\sum_i|\mathbf{p_i}
		\times \mathbf{\hat{n}}|}{\sum_i|\mathbf{p_i}|}\right)^2 ,
\end{align}
where, $ \hat{\mathbf{{n}}} $ represents the spherocity axis, which is the direction that minimizes the expression under parentheses, and $ \mathbf{p_i} $ denotes the three-momentum of the $ i^{\text{th}} $ particle in the final state. Simply put, the spherocity axis is the direction perpendicular to which the energy flow is minimized. By comparison, the thrust axis is defined as the direction along which the energy flow is maximum. This similarity between the thrust and spherocity axes is not surprising, and it has been established that at the one-loop level, the thrust and spherocity axes are identical. However, no well-defined method currently exists for computing the spherocity axis at higher loop levels.
In this study, we focus on processes with a three-body final state, for which the spherocity is given by the following expression~\cite{Georgi:1977sf, Ellis:1991qj}

\begin{align}\label{eq:defSphero}
	s_{p}(x_1,x_2,x_3)\, = \,\frac{16}{\pi^2} \left(\frac{(1-x_1)(1-x_3)(1-x_3)}{\text{max}(x_1^2, x_2^2, x_3^2)}\right),
\end{align}
where $ x_i $ are the energy fraction variables defined in eq.~(\ref{xi's}), 
for ease of notation we define a rescaled spherocity $ s(x_1,x_2,x_3) $ as
\begin{align}\label{eq:defRescSpheroc}
	s(x_1,x_2,x_3)\, = \,\frac{\pi^2}{16} \times  s_{p}\, = \,\frac{(1-x_1)(1-x_3)(1-x_3)}{\text{max}(x_1^2, x_2^2, x_3^2)}\,.
\end{align}
The above expression of spherocity shape variable consists of three equations based on which energy fraction is the largest. Thus, we have three different definitions of spherocity, just like thrust.
Let us first analyze the behavior of our event shape variable for different kinematical configurations.

\begin{figure}[h]
	\centering
	\includegraphics[width = 0.65\linewidth]{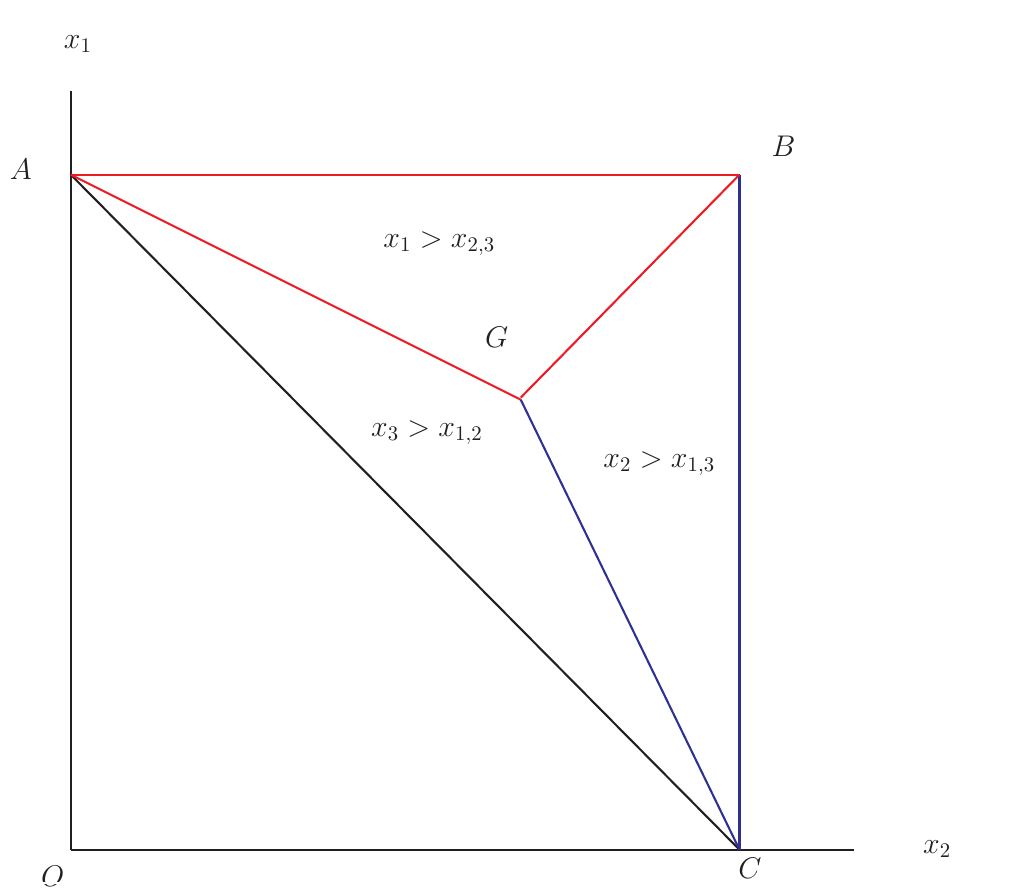}
	\caption{Dalitz plot for the process under consideration, the phase space is divided into three different parts based upon hierarchy of energy fraction variables. }
	\label{fig:SpheroDalitz}
\end{figure}

\begin{figure}[h]
	\centering \includegraphics[scale = 0.60]{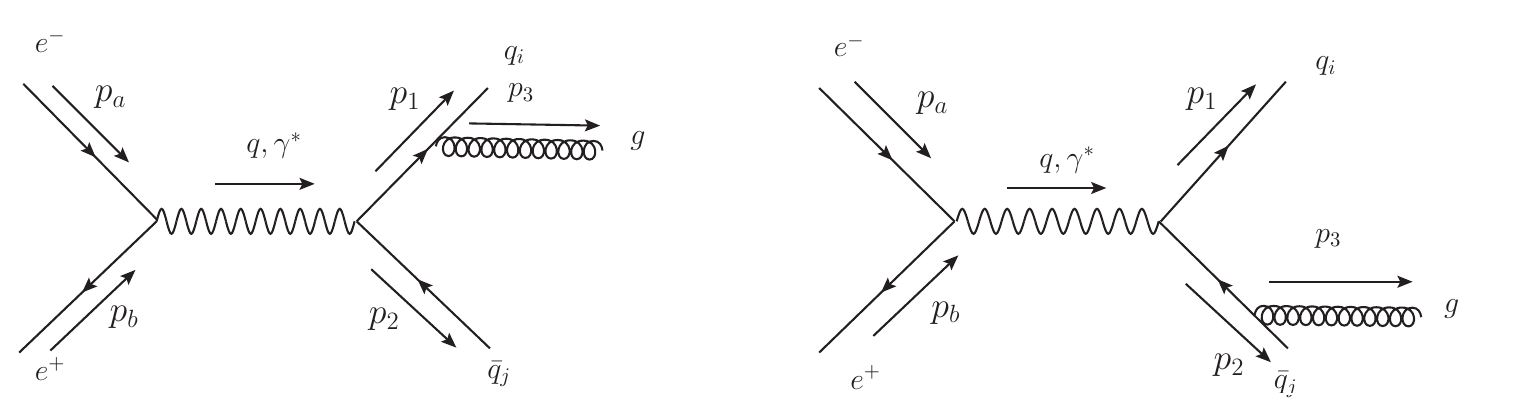}
	\caption{Feynman diagrams for the real emission of a gluon from the
		final state quark or anti-quark.}
	\label{processdiag}
\end{figure}

In figure~\ref{fig:SpheroDalitz}, the region around point $ B $ resembles a kinematical limit where $ x_1 \to 1, x_2 \to 1 $ and, $ x_3 \to 0 $ which corresponds to the emission of a soft gluon and back-to-back configuration between quark and anti-quark, for this configuration spherocity variable takes the value 0. Similarly, for the region around point $ A $, $ x_1 \to 1, x_2 \to 0 $ and, $ x_3 \to 1 $, which corresponds to a configuration where gluon is hard and is in back-to-back dijet configuration with quark, under this limit again $ s \to 0 $. Thus  $ s \to 0 $ for dijet event. In the case of a three-body final state, the isotropic shape is attained when $ x_1=x_2=x_3=2/3 $, i.e., all the particles have the same energies; for this configuration, our rescaled spherocity takes the value $ s=1/12\,(=0.0833) $ and it is the maximum value spherocity variable can take in our case of the three-body final state. Further about spherocity and how various geometries are treated within its definition can be found in~\cite{Brandt:1978zm}.
We aim to compute the leading contributions to the spherocity distribution at the next-to-leading order (NLO).
The leading order (LO) reaction at the parton level is
\begin{equation}
	e^{+}(p_b)+e^{-}(p_a)\rightarrow \gamma^*(q) \rightarrow q(p_1)  + {\overline q}(p_2)\,,
	\label{processeqnborn}
\end{equation}
where we assume all particles to be massless. For this case, $s \, = \, 0$.
At NLO the real emission process is
\begin{equation}
	e^{+}(p_b)+e^{-}(p_a)\rightarrow \gamma^*(q) \rightarrow q(p_1)  + {\overline q}(p_2) + g(p_3)\,.
	\label{processeqn}
\end{equation}
The diagrams for this process are shown in figure~\ref{processdiag}.
The limit $s = 0$ is approached when either $(i)$ the emitted gluon is soft
($p_3 \to 0$),  $(ii)$ the quark or the anti-quark is soft
($p_1\to 0 \ \text{or} \ p_2 \to 0$), or $(iii)$ any two final state
partons are collinear.
It is standard practice to define the dimensionless energy fractions for the
final state particles,
\begin{align}
	x_i&\, = \, \frac{2E_i}{Q} \quad \quad (i\, = \,1,2,3)\,,
\end{align}
where $Q$ is the total center of mass energy, with   $x_1+x_2+x_3\, = \,2$.
These variables follow the relations given below
\begin{eqnarray}
	(p_2+p_3)^2&\, = \,&2p_2\cdotp p_3\, = \,Q^2(1-x_1)\,,\nonumber  \\
	(p_1+p_3)^2&\, = \,&2p_1\cdotp p_3\, = \,Q^2(1-x_2)\,, \label{xi's} \\
	(p_1+p_2)^2&\, = \,&2p_1\cdotp p_2\, = \,Q^2(1-x_3)\nonumber\,.
\end{eqnarray}
The matrix element squared for the process in
eq.~(\ref{processeqn}) is
\begin{equation}\label{eq:MatrixElement}
	\overline{\sum} |\mathcal{M}(x_1,x_2)|^2\, = \,8(e^2e_q)^2 g_s^2 C_F N_c \frac{1}{3Q^2} \, \left[\frac{x_1^2+x_2^2}{(1-x_1)(1-x_2)}\right] \,,
\end{equation}
where $\alpha\, = \,\left( e^2/4 \pi\right)$, $e_q$ is the charge of quarks in the unit of fundamental electric charge $e$,
$\alpha_s\, = \, g_s^2/4\pi$,
$C_F \, = \,(N_c^2-1)/2N_c $, and $N_c$ is the number of quark colours. 
Having defined the shape variable and matrix elements squared for the process discussed above, we move to the computation of spherocity distribution at NLO in the next section. 
\section{Spherocity distribution at NLO} \label{sphero_NLO}

The spherocity distribution at NLO is given by
\begin{equation}	\label{eq:SP_dstr}
	\frac{d\sigma}{ds}\, = \,\frac{1}{2s_0}\int d\Phi_{3} \,\,\Bigg[\overline{\sum} \,|\mathcal{M}(x_1,x_2, x_3)|^2 \,\,  \delta\big(s-s(x_1,x_2,x_3)\big)\Bigg]\, ,
\end{equation}
where $ s_0 $ is center of mass energy. The matrix elements squared $ 	\overline{\sum} |\mathcal{M}(x_1,x_2)|^2\, $ and the spherocity shape variable $ s(x_1, x_2, x_3) $ are given in eqs.~\eqref{eq:MatrixElement} and~\eqref{eq:defRescSpheroc} respectively in terms of energy fraction variables. 
The Lorentz invariant phase space factor for the three-body final
state is
\begin{align}
	d \Phi_3 \, = \,& \frac{d^3\mathbf{p_1}}{(2\pi)^32E_1}\frac{d^3\mathbf{p_2}}{(2\pi)^32E_2}\frac{d^3\mathbf{p_3}}{(2\pi)^32E_3}(2\pi)^4 \delta^4(q-p_1-p_2-p_3)\,.
	\label{3bfull} 
\end{align}
As $\overline{\sum} |\mathcal{M}(x_1,x_2)|^2$ depends only on
the energy fractions, we write the phase space factor in terms of energy fractions as well, moving ahead in eq.~\eqref{3bfull} one can trivially integrate over $\bf p_{3}$
followed by an integration over  $\bf{p_{1}}$ and $\bf{p_{2}}$ and obtain
the three-particle phase space measure as
\begin{align}
	d \Phi_3\, = \, &\frac{Q^2}{16(2\pi)^3} \, dx_1 \, dx_2 \,  dx_3 \,\, \delta  \Big(x_1+x_2+x_3-2\Big).
	\label{3bx1x2}
\end{align}
The phase space region of eq.~(\ref{3bx1x2}) is shown in figure~\ref{fig:SpheroDalitz} in form of a plot between $ x_1 \,  \text{and}\,  x_2 $, with every point in the plane satisfying the constraint $x_1+x_2+x_3\, = \,2$. 
Upon substituting the above expressions in eq.~\eqref{eq:SP_dstr} spherocity distribution takes the form
\begin{align} \label{eq:MainFormula}
	\frac{1}{\sigma_0(s_0)}\frac{d\sigma}{ds}\, = \,\frac{2\alpha_s}{3\pi}\int
	_{0}^{1} dx_1 \int_{0}^{1-x_1} dx_2 \,\,
	\Bigg[\frac{x_1^2+x_2^2}{(1-x_1)(1-x_2)}
	& \,\, \delta \left(s-\frac{(1-x_1)(1-x_3)(1-x_3)}{\text{max}(x_1^2, x_2^2, x_3^2)}\right)\Bigg]\,,
\end{align}
where $x_{3}\, = \, 2 -x_{1} -x_{2}$, and $\sigma_0(s_0)$ is the LO
cross-section, given by
\begin{equation}
	\sigma_0(s_0)\, = \,\frac{4\pi\alpha^2  e_q^2\, N_c}{3s_0}\,,
	\label{born cross section}
\end{equation}
%
For our purposes, we label the contributions from three distinct 
in the phase space integration in eq.~(\ref{eq:MainFormula}), defined
by either $x_1$, $x_2$, and $x_3$ being the largest. The division of phase space region shown in figure~\ref{fig:SpheroDalitz} is for three particles ($ q, \, \bar{q}, \,\text{and} \, g  $) in the final state. The regions ABG, AGC, and BGC are the unconstrained phase space, these regions are divided on the basis of hierarchy of energy fraction. 
Computation of the event shape distribution necessitates imposing constraints on the phase space accessible to final state particles. These constraints are implemented through a $ \delta $-function embedded between the matrix elements and the phase space factor, as given in eq.~(\ref{eq:SP_dstr}). The specific nature of this constraint depends on the chosen shape variable and its numerical value. In section~\ref{sec:PHASEspacePLOTS}, we delve into the constraints imposed by spherocity, utilizing multiple phase space plots at varying spherocity values for detailed illustration.

To classify the contributions to the cross-section distribution, we categorize the constrained phase space into three distinct regions based on the definitions associated with the spherocity shape variable. We label these regions as
\begin{itemize}
	\item[(i)]{Region-I\,:}  First definition of spherocity where $ x_1^2 $ is largest
	\item[(ii)]{Region-II\,:}  Second definition of spherocity where $ x_2^2 $ is largest 
	\item[(iii)]{Region-III\,:} Third definition of spherocity where $ x_3^2 $ is largest
\end{itemize}
We are now ready to calculate the spherocity distribution. In the following subsections, we will compute this distribution by analyzing the contributions from three distinct regions: regions-I, II, and III.

\subsection{Contribution from region-I}\label{section:region1}

\begin{figure}[h]
	\centering
	\includegraphics[width = 0.50\linewidth]{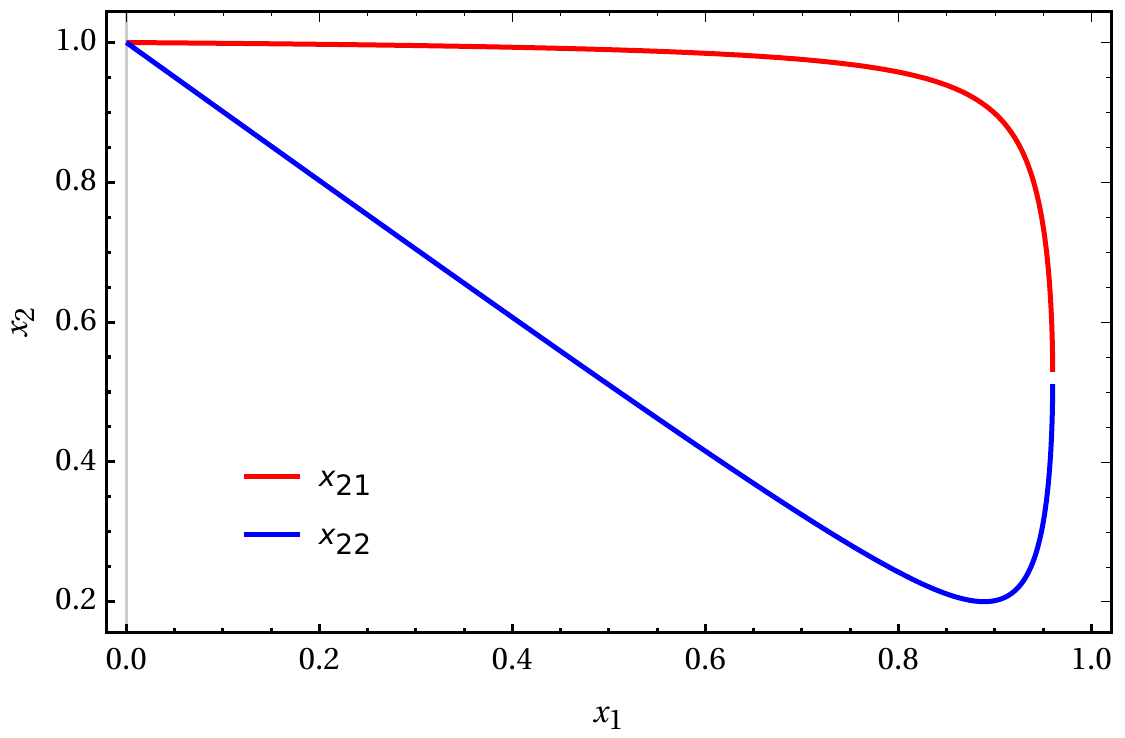}
	\caption{Plot for $ x_2 $ against $ x_1 $ as constrained by the region-I definition of spherocity in eq.~\eqref{eq:ExactR1} at $ s=0.01$.}
	\label{fig:rootR1}
\end{figure}

Firstly, we start with the computation of spherocity distribution from region-I. Remember that during the computation of shape distribution from different regions (I, II, and III) of phase space,  only the definition of shape variables will change; the expressions of phase space factor and matrix elements remain the same. 
Plugging the expressions of matrix element and shape variable given in eqs.~\eqref{eq:MatrixElement} and~\eqref{eq:defRescSpheroc}, the spherocity distribution from region-I takes the following  form

\begin{align}\label{eq:ExactR1}
	\hspace{-0.4cm}	\frac{1}{\sigma_0(s_0)} \frac{d \sigma}{d s} {\Bigg \vert_{\text{I}}}\, = \,\frac{2\alpha_s}{3\pi}  \int _{0}^{1} dx_2 \int_{0}^{1-x_2} dx_1 \,  \Bigg[\frac{x_1^2+x_2^2}{(1-x_1)(1-x_2)} \,\, \delta\left(s- \frac{(1-x_1)(1-x_3)(1-x_3)}{x_1^2}  \right)\Bigg]\,,
\end{align}
firstly, we perform the integration over $ x_2 $ with the aid of $ \delta $-function embedded between phase space factor and matrix element. This $ \delta $-function constrains the allowed phase space for the particles present in the final state by restricting the values these energy fractions can take; the value of $ s $ regulates the restrictions. The $ \delta $-function in above expression has two roots of $ x_2 $ and their respective expressions in terms of $ x_1 $ and $ s $ is given below

\begin{align}\label{eq:Exact-deltaroots-R1}
	x_{21}(x_1,s)&\, = \, \frac{2+x_1\left( -3+x_1-\sqrt{-1+x_1}\sqrt{-1+4s+x_1} \,  \right)}{2-2x_1} \nonumber \, , \\
	x_{22}(x_1,s)&\, = \, \frac{2+x_1\left( -3+x_1+\sqrt{-1+x_1}\sqrt{-1+4s+x_1} \,  \right)}{2-2x_1}\, .
\end{align}
In the dijet limit ($ s\to 0 $), the above roots approximate to 
\begin{align}
	x_{21}& \to 0  \quad ,  \quad
	x_{22} \to  1-x_1 \, .
\end{align}
To visualize the constraint established by the $ \delta $-function in eq.~\eqref{eq:ExactR1}, we show a plot in figure~\ref{fig:rootR1}, where the graphical representation of these roots given in eq.~\eqref{eq:Exact-deltaroots-R1} are displayed as a function of $ x_1 $ at $ s=0.01 $. 
The plot shows that the allowed phase space region remains quite large at this specific value of $s$. As a result, the leading contributions to the distribution will primarily originate from this part of phase space.
To proceed further we perform the $ x_2 $ integration by substituting the first root ($ x_{21} $) in eq.~\eqref{eq:ExactR1}, after performing the integration, the integrand for $ x_1 $ takes the form

\begin{align}\label{eq:R1I1}
	&\frac{1}{\sigma_0(s_0)}  \frac{d \sigma}{d s} {\Bigg \vert_{\text{I}_1}}\, = \, 
	\frac{2\alpha_s}{3\pi}  \int _{s}^{1-4s} dx_1  \nonumber  \\	& \times \left[\frac{x_1 \left(x_1 \left(\sqrt{x_1-1} \sqrt{4 s+x_1-1}-2 s-3 x_1+5\right)-2 \left(\sqrt{x_1-1} \sqrt{4 s+x_1-1}+2\right)\right)+2}{(x_1-1)^2 \left(-\sqrt{x_1-1} \sqrt{4 s+x_1-1}+4 s+x_1-1\right)} \right]\, ,
\end{align}
similar substitution of second root ($x_{22}  $) in eq.~\eqref{eq:ExactR1} yields

\begin{align}\label{eq:R1I2}
	&\frac{1}{\sigma_0(s_0)}  \frac{d \sigma}{d s} {\Bigg \vert_{\text{I}_2}}\, = \, 
	\frac{2\alpha_s}{3\pi}  \int _{s}^{1-4s} dx_1 \,  \nonumber \\	& \times \left[ \frac{x_1 \left(-2 \sqrt{x_1-1} \sqrt{4 s+x_1-1}+x_1 \left(\sqrt{x_1-1} \sqrt{4 s+x_1-1}+2 s+3 x_1-5\right)+4\right)-2}{(x_1-1)^2 \left(\sqrt{x_1-1} \sqrt{4 s+x_1-1}+4 s+x_1-1\right)}	  \right] \, , 
\end{align}
adding the above two expressions the final integrand over $ x_1 $ takes the following form
\begin{align}\label{eq:R1-Integrand}
	\frac{1}{\sigma_0(s_0)}  \frac{d \sigma}{d s} {\Bigg \vert_{\text{I}}} \, = \, \frac{2\alpha_s}{3\pi} \int _{s}^{1-4s} dx_1 \, \left[\frac{x_1 \left(s (x_1-4)-x_1^2+x_1-1\right)+1}{s (x_1-1)^{3/2} \sqrt{4 s+x_1-1}} \right]\,.
\end{align}
Combining the two individual integrands in eqs.~\eqref{eq:R1-Integrand} and \eqref{eq:R1I2} resulted in a more compact final form, as shown in eq. \eqref{eq:R1-Integrand}. Integrating this expression over $x_1$ with appropriate limits, e.g., $s$ and $1-4s$ for lower and upper bounds, respectively, yields logarithmic terms as a function of $s$. This behavior can be anticipated from the integrand's denominator. The term $(x_1-1)^{3/2}$ exhibit divergent logarithmic behavior as $x_1$ approaches 1, similarly, the term $(4s+x_1-1)^{1/2}$ displays analogous behavior when $s$ tends to 0 and $x_1$ approaches 1.
In this context, the upper limit ($1-4s$) reflects the configuration where a quark-antiquark pair is emitted in a back-to-back configuration, accompanied by an emission of soft gluon, and the lower limit ($ s $) corresponds to quark having having vanishing momenta and anti-quark, and gluon in a back-to-back configuration. This soft gluon emission can occur at a wide angle or collinearly with the quark's momentum axis, the soft collinear configuration contributes most significantly to the overall differential cross-section distribution.

The expression for spherocity distribution at NLO from region-I has the following form
\begin{align}\label{eq:R1NLO}
	\frac{1}{\sigma_0(s_0)}  \frac{d \sigma}{d s} {\Bigg \vert_{\text{I}}} \, = \,  \alspi &
	\Bigg[\frac{-2 \left(4 s^2+1\right) \log 4 s+2s \log 4 s}{s}- \frac{(s (7 s-10)+6) \sqrt{\frac{4}{s-1}+5}}{2s}  \nonumber \\
	&\hspace{3.3cm}+\frac{4 (s (4 s-1)+1) \log \left(\sqrt{1-s}+\sqrt{1-5 s}\right)}{s}\Bigg] \,.
\end{align}
The presence of $ s^{-1} $ and $ \log s $ terms indicate the diverging behavior of spherocity distribution in the dijet limit as expected. 
In order to examine the dijet behavior of this distribution, we expand eq.~\eqref{eq:R1NLO} around the limit $ s \, = \,0 $ as

\begin{align}
	\frac{1}{\sigma_0(s_0)}  \frac{d \sigma}{d s} {\Bigg \vert_{(\text{NLO+NNLP})_I}} \, = \, \alspi	\left(\frac{-3-2\log s}{s}+5+2\log s+ \left(-\frac{13}{2}-8\log s\right)s + \mathcal{O}(s^2)\right)\, ,
\end{align}
The first term in the brackets is the LP term (in $ s $), consisting of the LL and NLL. The LL term at LP arises from soft and collinear gluon emission, while the NLL term at LP originates from hard collinear gluon emission. The subsequent set of terms proportional to $ s^0 $ constitutes the NLP term. The NLP term also consists of LL and NLL. LL at NLP receives contributions from many different configurations, i.e., next-to-soft gluon emissions and soft fermionic emissions and others, the exact sources are not entirely understood, and it is an exciting task at hand. In one of our previous studies~\cite{Agarwal:2023fdk}, we attempted to calculate these LL terms at NLP for thrust and $ c $-parameter distributions using approaches of shifted kinematics~\cite{DelDuca:2017twk} and soft fermionic emissions~\cite{vanBeekveld:2019prq, vanBeekveld:2019lwy}.  
Further terms proportional to $s$ and higher powers of $s$ represent next-to-next-to-leading power (NNLP) accuracy and beyond.
The logarithms in the above expression diverge at the threshold (dijet limit), leading to diverging results in fixed-order calculations.
In these limiting cases, where logarithmic terms arise, resummation to all orders becomes crucial. This technique effectively sums up an infinite series of corrections, ensuring reliable and physically meaningful results. Therefore, computing these logarithmic corrections (LLs and NLLs) is essential for obtaining theoretically sound predictions.
Up to NLO+NLP, spherocity distribution from region-I reads as 
\begin{align}\label{eq:R1NLP}
	\frac{1}{\sigma_0(s_0)}  \frac{d \sigma}{d s} {\Bigg \vert_{(\text{NLO+NLP})_I}} \, = \, \alspi	\left(\frac{-3-2\log s}{s}+5+2\log s + \mathcal{O}(s)\right) \, .
\end{align}
Thus, as the kinematical phase space available for particle interactions becomes smaller, the cross-section distribution is increasingly dominated by contributions arising from logarithmic factors.

Next, we shift our attention to computing spherocity distribution from region-II. The definition of shape variable from region-II has the form 
\begin{align}\label{eq:S2}
	s(x_1, x_2, x_3)\,  = \,\frac{(1-x_1)(1-x_3)(1-x_3)}{x_2^2}\,.
\end{align}
As the phase space integration and matrix elements in eq.~\eqref{eq:MainFormula} treat quarks and anti-quarks identically, the contributions from region-II are precisely the same as those from region-I.
Based upon the argument given above, the contribution to spherocity distribution from region-II can directly be written as
\begin{align}\label{eq:R2NLP}
	\frac{1}{\sigma_0(s_0)}  \frac{d \sigma}{d s} {\Bigg \vert_{(\text{NLO+NLP})_{II}}} \, = & \, \alspi	\left(\frac{-3-2\log s}{s}+5+2\log s + \mathcal{O}(s)\right) \,.  \nonumber \\ = & \,	\frac{1}{\sigma_0(s_0)}  \frac{d \sigma}{d s} {\Bigg \vert_{(\text{NLO+NLP})_I}} \, 
\end{align}
Having computed the contributions to the spherocity distribution from regions-I and II, we observe that regions-I and II contribute towards the LL and NLL of LP and NLP. Moving ahead, we will now focus on the computation of spherocity distribution from region-III.

\subsection{Contributions from region-III } \label{section:region3}

In region-III, the spherocity shape variable takes the following form


\begin{align}\label{eq:S3}
	s(x_1, x_2, x_3)\,  = \,\frac{(1-x_1)(1-x_3)(1-x_3)}{x_3^2}\,.
\end{align}
Plugging the expressions of matrix element squared from eq.~\eqref{eq:MatrixElement} and event shape from above expression into eq.~\eqref{eq:SP_dstr}, the distribution takes the following form

\begin{align}\label{eq:ExactR3}
	\hspace{-0.4cm}	\frac{1}{\sigma_0(s_0)} \frac{d \sigma}{d s} {\Bigg \vert_{\text{III}}}\, = \,\frac{2\alpha_s}{3\pi}  \int _{0}^{1} dx_3  \int_{0}^{1-x_3} & dx_2  \,  \Bigg[\frac{(2-x_2-x_3)^2+x_2^2}{(x_2+x_3-1)(1-x_2)} \,\, \nonumber \\
	\times \, & \delta\left(s- \frac{(x_2+x_3-1)(1-x_2)(1-x_3)}{x_3^2}  \right)\Bigg]\,,
\end{align}
here, we chose to substitute $ x_1 =2-x_2-x_3 $ in expressions of matrix element, shape variable, and phase space factor to facilitate integrations.
Following the similar steps as in section~\ref{section:region1}, we perform $ x_2 $ integration first followed by $ x_3 $ integration, the $ \delta $-function roots for $ x_2 $ in terms of $ x_3 $ and $ s $ have the following form:
\begin{align}\label{eq:R3roots}
	x_{21}(x_3,s)&=1+\frac{x_3}{2} \left(\frac{\sqrt{4 s+x_3-1}}{\sqrt{x_3-1}}-1\right), \nonumber \\
	x_{22}(x_3,s)&=1-\frac{x_3}{2} \left(\frac{\sqrt{4 s+x_3-1}}{\sqrt{x_3-1}}+1\right).
\end{align}
The behavior  of above  roots in limit $ s \to 0 $ is
\begin{align}
	x_{21} \to 1\,, \quad x_{22} \to 1-x_3\,,
\end{align}
\begin{figure}[h]
	\centering
	\includegraphics[width = 0.52\linewidth]{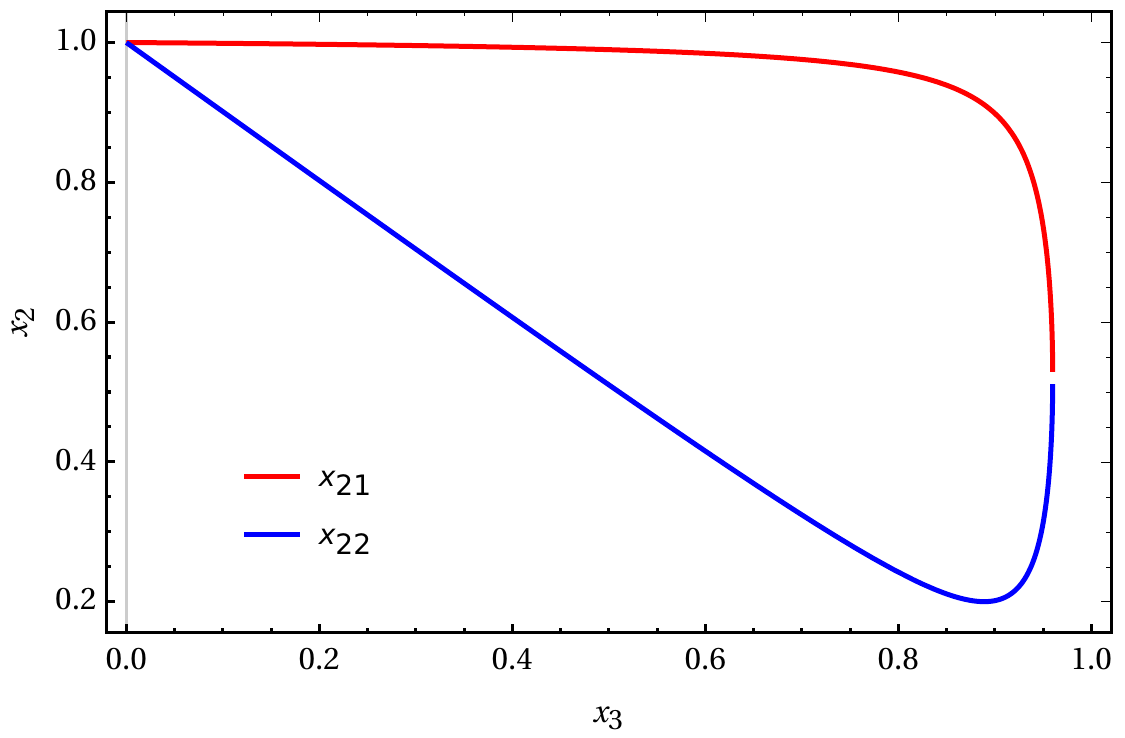}
	\caption{Plot for $ x_2 $ against $ x_3 $ as constrained by the region-III definition of spherocity in eq.~\eqref{eq:ExactR3} at $ s=0.01$.}
	\label{fig:rootR3}
\end{figure}
In order to visualize clearly, we plot the roots given in eq.~\eqref{eq:R3roots} in figure~\ref{fig:rootR3}. We observe that the constraints imposed on the allowed phase space by the definition of spherocity in region-III do not significantly reduce the accessible phase space. Furthermore, the region around $x_3=0$ remains accessible, corresponding to the soft gluon region as previously discussed in figure~\ref{fig:SpheroDalitz}, and we expect to receive leading contributions at LP and NLP from this region.
Substituting the first root of $ x_2 $ in eq.~\eqref{eq:ExactR3}  we get
\begin{align}  \label{eq:R3I1}
	\frac{1}{\sigma_0(s_0)}  \frac{d \sigma}{d s} {\Bigg \vert_{\text{III}_1}}\, = \, 
	\frac{2\alpha_s}{3\pi}  \int _{s}^{1-4s} dx_3 \left[\frac{x_3 (x_3 (2 s+x_3-3)+4)-2}{s \sqrt{x_3-1} x_3 \sqrt{4 s+x_3-1}}\right] \, ,
\end{align}
similarly upon substitution of second root we get
\begin{align}\label{eq:R3I2}
	\frac{1}{\sigma_0(s_0)}  \frac{d \sigma}{d s} {\Bigg \vert_{\text{III}_2}}\, = \, 
	\frac{2\alpha_s}{3\pi}  \int _{s}^{1-4s} dx_3 \left[\frac{x_3 (x_3 (2 s+x_3-3)+4)-2}{s \sqrt{x_3-1} x_3 \sqrt{4 s+x_3-1}}\right] \, ,
\end{align}
above two expressions are identical, adding them the final integral attains the following form

\begin{align}\label{eq:R3I}
	\frac{1}{\sigma_0(s_0)}  \frac{d \sigma}{d s} {\Bigg \vert_{\text{III}}}\, = \, 
	\frac{2\alpha_s}{3\pi}  \int _{s}^{1-4s} dx_3 \left[\frac{2x_3 (x_3 (2 s+x_3-3)+4)-4}{s \sqrt{x_3-1} x_3 \sqrt{4 s+x_3-1}}\right] \, .
\end{align}
Similar to eq.~\eqref{eq:R1-Integrand}, the integrand in the above expression contains terms $ (x_3-1  )^{1/2}$ and $ (4s+x_3-1 )^{1/2} $ in the denominator, due to which there is a diverging behavior around $ x_3 =1 $ in the dijet limit. Interestingly, the denominator also contains a factor of $ x_3 $, which is divergent at $ x_3=0 $. Thus, the soft gluon contribution will also be captured from this integrand.
Performing the phase space integration over the final variable $ x_3 $, the spherocity distribution from region-III definition takes the following form

\begin{align}\label{eq:R3NLO}
	\frac{1}{\sigma_0(s_0)}  \frac{d \sigma}{d s} {\Bigg \vert_{\text{III}}}\, = \,	\alspi & \Bigg[\frac{8 \tanh ^{-1}\left(\sqrt{\frac{1-5 s}{4 s^2-5 s+1}}\right)}{\sqrt{1-4 s} s}+\log \left(\frac{256 s^8}{\left(\sqrt{5 s^2-6 s+1}-3 s+1\right)^8}\right) \nonumber \\
	&+\frac{\left(s^2+1\right) \log \left(\frac{16 s^4}{\left(\sqrt{5 s^2-6 s+1}-3 s+1\right)^4}\right)}{s}+\frac{(-s-3) \sqrt{(s-1) (5 s-1)}}{s}\Bigg] \,.
\end{align}
Expanding the above expression around $ s=0 $, we get
\textcolor{black}{ \begin{align}\label{eq:R3NLOser}
		\frac{1}{\sigma_0(s_0)}  \frac{d \sigma}{d s} {\Bigg \vert_{\text{III}}}\, = \,	\alspi \left( \frac{-3-4\log s}{s} -8\log s +(-27-44\log s)s +\mathcal{O}(s^2) \right)\,.
\end{align}}
The above expression shows that the region-III contributes substantially at NLO. As one can notice, the LLs at LP and NLP both receive contributions from this region, which is quite interesting and is in contrast to the thrust distribution~\cite{Agarwal:2023fdk} where we did not receive contribution to LL at LP from region-III. We study this aspect of spherocity distribution in detail later in section~\ref{sec:PHASEspacePLOTS}.
Combining the individual expressions computed from the three regions eqs.~\eqref{eq:R1NLO},~\eqref{eq:R2NLP} and~\eqref{eq:R3NLO},   the final form of spherocity distribution at NLO for the process in  eq.~\eqref{processeqn} is
\begin{align}\label{eq:NLO}
	\frac{1}{\sigma_{0}(s_0)} \frac{d \sigma}{d s} \Bigg\vert_{\text{NLO}}  \, = \,  \, \frac{2 \alpha_{s}}{3 \pi} &
	\Bigg[\frac{-4 \left(4 s^2+1\right) \log 4 s+2s \log 4 s}{s}- \frac{(s (7 s-10)+6) \sqrt{\frac{4}{s-1}+5}}{s}  \nonumber \\
	&+\frac{8 (s (4 s-1)+1) \log \left(\sqrt{1-s}+\sqrt{1-5 s}\right)}{s} + \frac{8 \tanh ^{-1}\left(\sqrt{\frac{1-5 s}{4 s^2-5 s+1}}\right)}{s\sqrt{1-4 s}}\nonumber \\
	&+8 \log \left(\frac{2 s}{\sqrt{5 s^2-6 s+1}-3 s+1}\right)
	-\left(\frac{s+3}{s}\right) \sqrt{(s-1) (5 s-1)} \\
	&   +\left(\frac{4s^2+4}{s}\right) \log \left(\frac{2s}{\sqrt{5 s^2-6 s+1}-3 s+1}\right)
	\Bigg] \nonumber \,.
\end{align}
The expression in eq.~\eqref{eq:NLO} contains all the leading corrections of spherocity distribution from all the possible geometrical configuration of the particles in final state. 
Expanding this NLO result  around $ s = 0 $, spherocity distribution up to NNLP reads as 

\begin{align}\label{eq:NLOfinal}
	\frac{1}{\sigma_{0}(s_0)} \frac{d\sigma}{d s} \Bigg\vert_{\text{NLO+NNLP}}  \, = \,  \, \frac{2 \alpha_{s}}{3 \pi} \left( \frac{-9-8\log s}{s}+10-4\log s - 20\left(2+3\log s\right)s +\mathcal{O}(s^2) \right)\, .
\end{align}
In the above expression, we have displayed results up to NNLP only. However, higher power terms can be reproduced from eq.~\eqref{eq:NLO}.
These results, especially the one obtained from eq.~\eqref{eq:ExactR3}, are unique in that the third definition of spherocity can also capture the LL at LP opposite to thrust.
This marks the complete computation of spherocity distribution at NLO for the process in eq.~\eqref{processeqn}.

\section{Phase space constraints and numerical assessment}\label{sec:PHASEspacePLOTS}
This section analyzes the spherocity distribution computed in eq.~\eqref{eq:R3NLOser}. The presence of LL at the LP from eq.~\eqref{eq:ExactR3} is a fascinating and distinctive feature of the entire distribution. In order to gain a proper understanding of this phenomenon,
Firstly, we examine the constraints imposed on the phase space by the definition of spherocity, and we plot the phase space subjected to these constraints. In the following subsection, we plot the spherocity distribution as computed from regions-I, II, and III.
\begin{figure}[h]
	\centering
	\includegraphics[width = 0.52\linewidth]{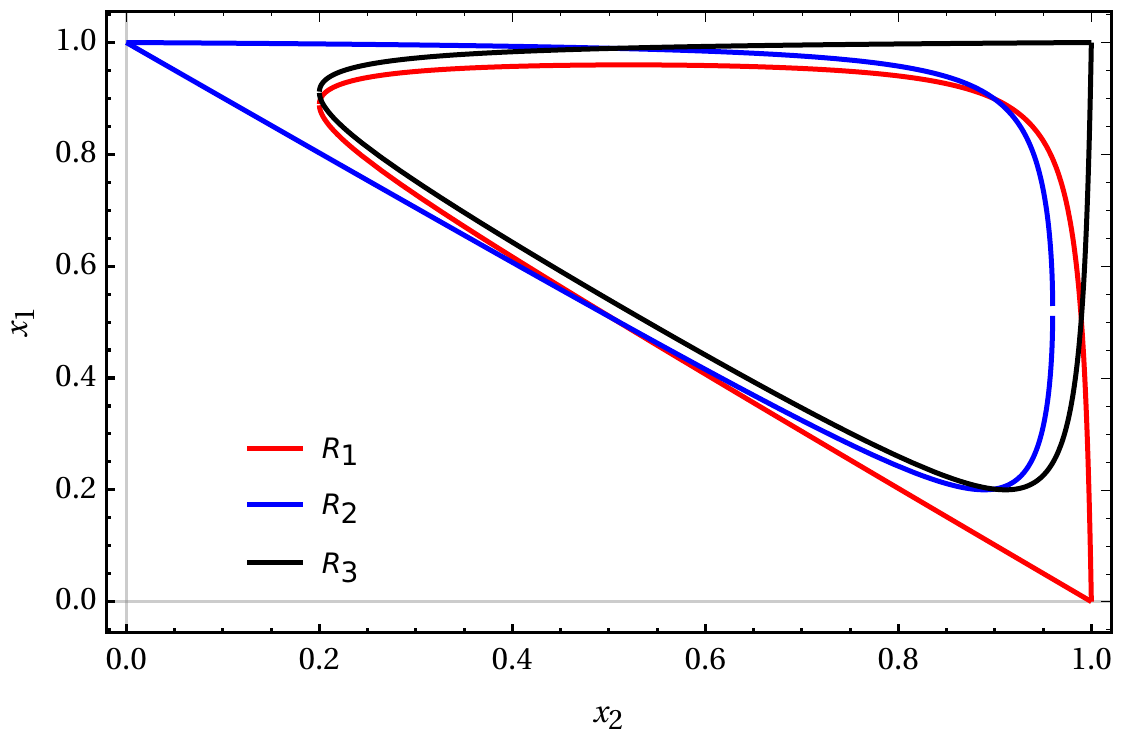}
	\caption{Plot for $ x_1 $ against $ x_2 $ at $ s=0.01 $ for  three spherocities  }
	\label{fig:x1R1-R3@s=0.01}
\end{figure}

\subsection{ Phase space analysis}
Here, we analyze the kinematics of the phase space to investigate the effects of three distinct definitions of spherocity, from eq.~\eqref{eq:defRescSpheroc}. These definitions are based on which particle in the final state has the largest energy fraction.
\begin{figure}[]
	\centering
	\vspace{-3mm}
	\hspace{-0.2cm}
	\subfloat[]{\hspace{0.3cm}\includegraphics[scale=0.36]{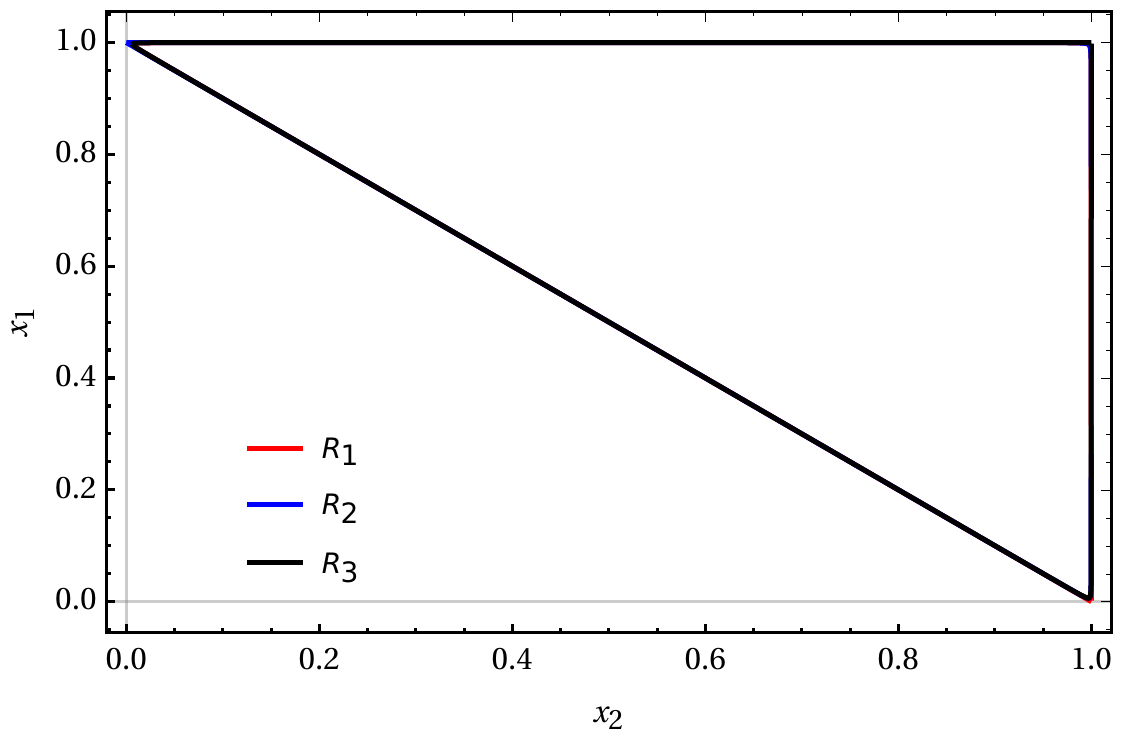} } 
	$ \; $
	\subfloat[]{\hspace{0.3cm}\includegraphics[scale=0.36]{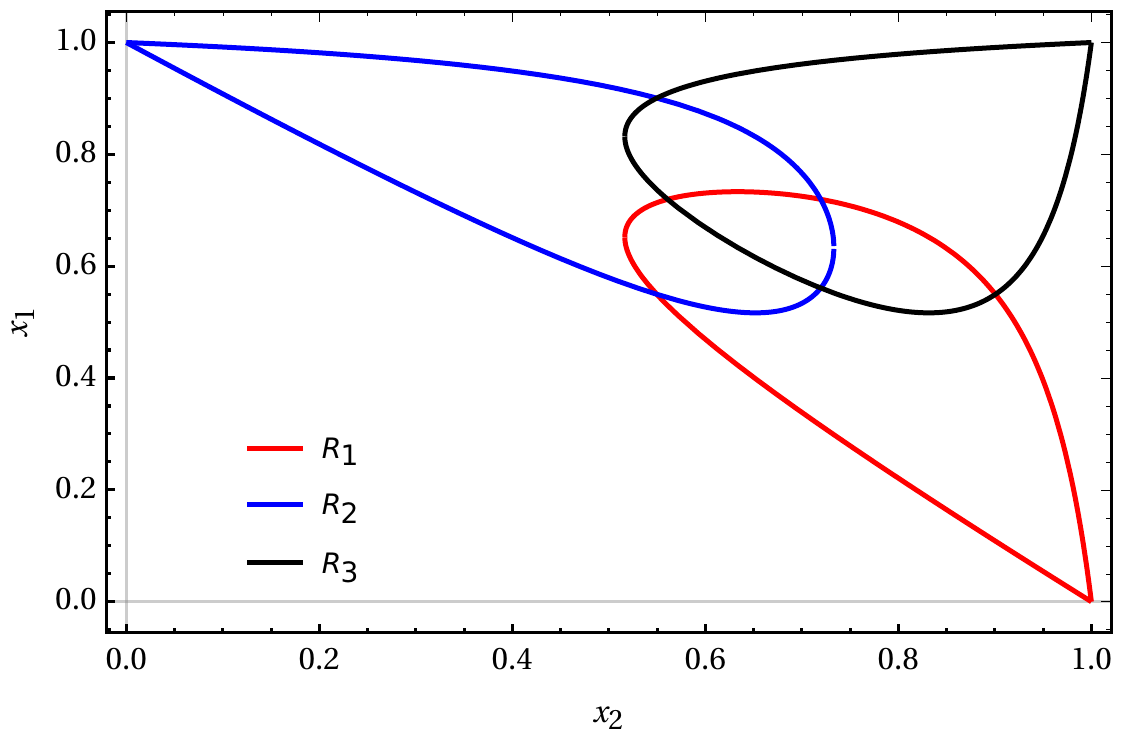} } 
	$ \; $
	\caption{Plot for $x_1$ against $x_2$  (a) s$= 10^{-5} $, (b) s $= 1/15 $, for  three different regions}
	\label{fig:x1-x2_extraPLOTS}
\end{figure}   
When solving the $\delta$-function in eq.~\eqref{eq:ExactR1}, the $ \delta $-function imposes constraints on the phase space by limiting the values these energy fractions can take. Figure~\ref{fig:x1R1-R3@s=0.01}, shows the plot of $x_1$ against $x_2$ at $s=0.01$, the red curve represents the variation of $x_1$ and $x_2$ as restricted by the first definition of spherocity from eq.~\eqref{eq:ExactR1}, where $x_1$ is considered the largest. Similarly, the blue curve shows the variation when $x_2$ is largest, and the black curve shows the variation for the definition where $x_3$ is considered largest.
The most striking feature of this plot is the black curve. It extends towards the top-right corner (point  B  in figure~\ref{fig:SpheroDalitz}), reaching deeper into this region than the red and blue curves. This specific corner holds significance because it represents a kinematical limit where $ x_1 $ and $  x_2 $ approach 1, forcing $ x_3 \to 0 $. In geometrical terms, this limit corresponds to a back-to-back configuration of a quark and anti-quark, along with an emission of a soft gluon. The closer we reach this corner, the more we can capture the leading corrections from the shape distribution.  
Contrary to thrust distribution~\cite{Agarwal:2023fdk}, region-III also captures the  LL at the LP in the spherocity distribution. These three curves completely overlap with each other when $s$ is taken very small ($s=10^{-5}$), as shown in figure~\ref{fig:x1-x2_extraPLOTS}\textcolor{blue}{a}.

\begin{figure}[]
	\centering
	\vspace{-3mm}
	\hspace{-0.2cm}
	\subfloat[]{\hspace{0.3cm}\includegraphics[scale=0.36]{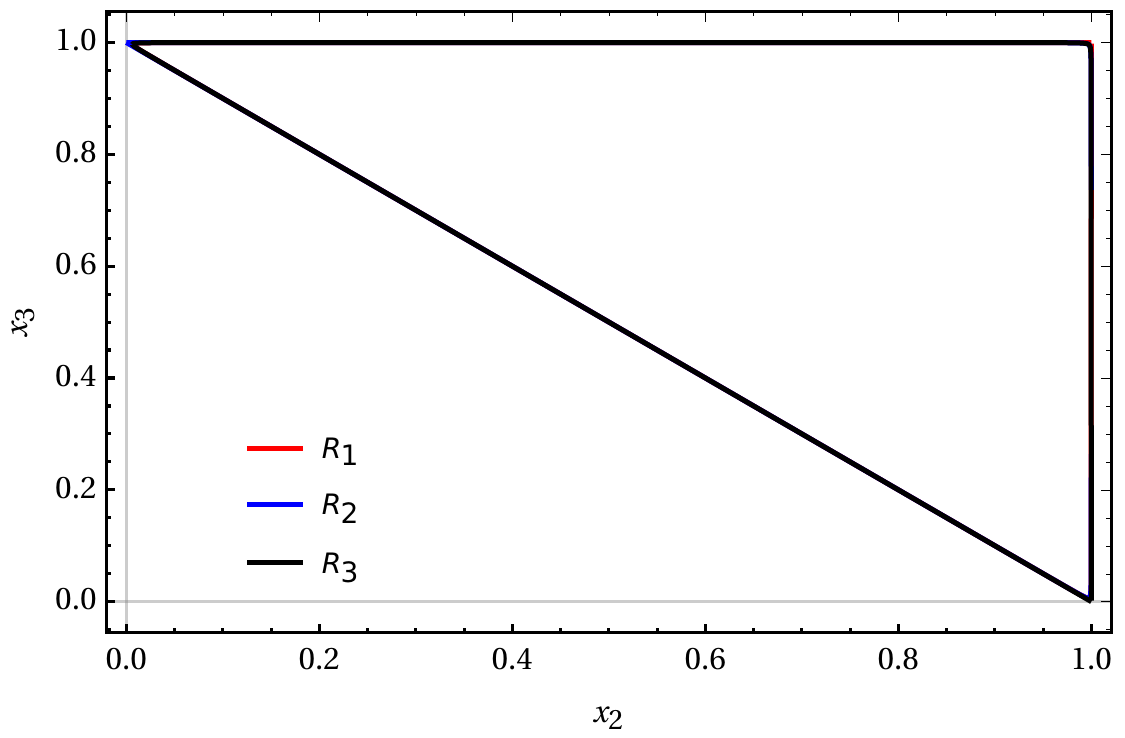} } 
	$ \; $
	\subfloat[]{\hspace{0.3cm}\includegraphics[scale=0.36]{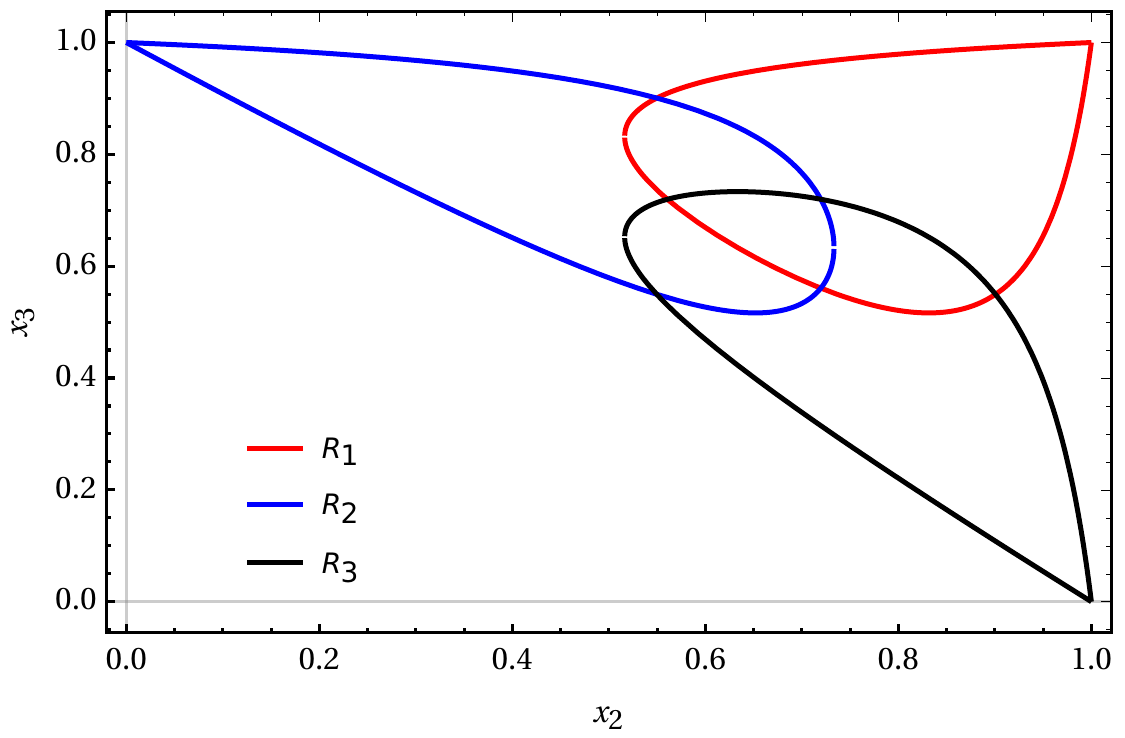} } 
	$ \; $
	\caption{Plot for $x_3$ against $x_2$ at (a) s$= 10^{-5} $  and (b) s $= 1/15 $, for three different regions}
	\label{fig:x3-x2_extraPLOTS}
\end{figure}

To further dissect this, we plot $ x_1 $ as a function of $ x_2 $ at $ s=1/15 $ in Figure~\ref{fig:x1-x2_extraPLOTS}\textcolor{blue}{b}, subject to their respective restrictions. Here, we notice that these curves decouple from each other and shift towards different edges of phase space, depending on the definition of the shape variable being used. To be precise,     figures~\ref{fig:x1-x2_extraPLOTS}\textcolor{blue}{b} and \ref{fig:x3-x2_extraPLOTS}\textcolor{blue}{b} show that the individual plots shift towards the edge where $ x_i $ is smallest, for example black (blue) curve shifts towards the edge where $ x_3 $ ($ x_2 $) is smallest, and red curve shifts towards the bottom-right edge where $ x_1 $ is smallest.  

\begin{figure}[]
	\centering
	\includegraphics[width = 0.52\linewidth]{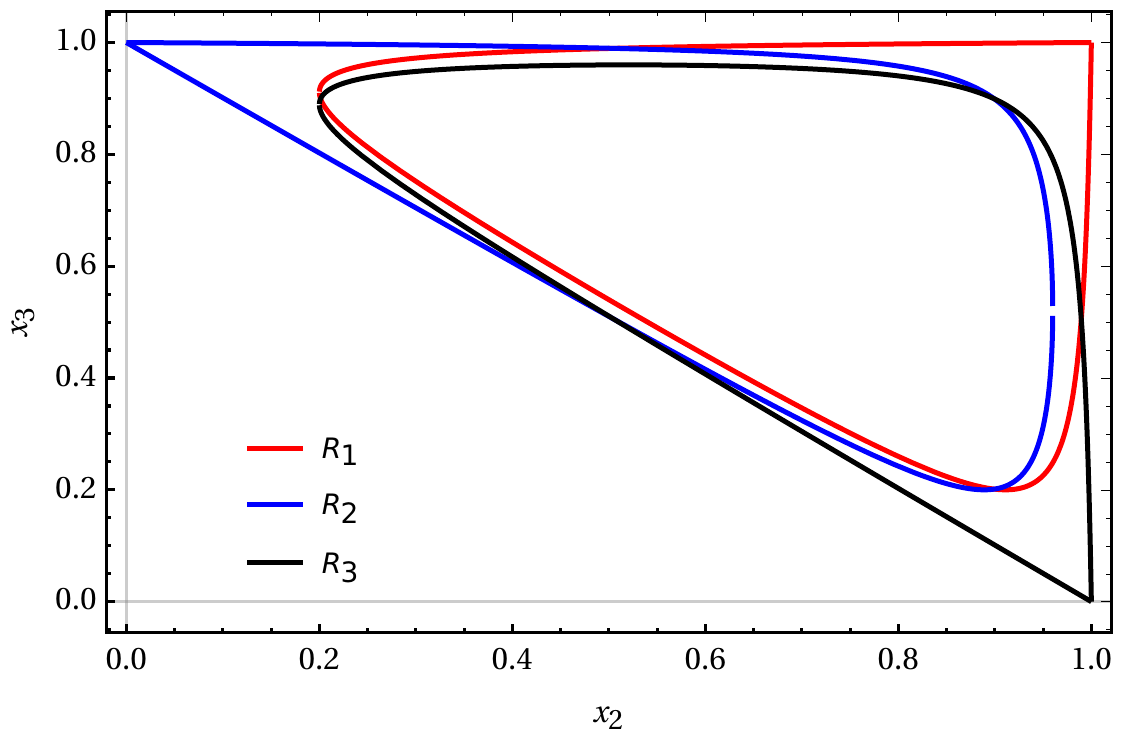}
	\caption{Plot for $ x_3$ against $ x_2 $ at $ s=0.01 $ for  three spherocities  }
	\label{fig:x3R1-R3}
\end{figure}

The constraints imposed by event shape thrust differ from those observed in the spherocity plots shown in figures~\ref{fig:x1R1-R3@s=0.01} to~\ref{fig:x3-x2_extraPLOTS}.
Figure~\ref{fig:SpheroDalitz} depicts the unconstrained phase space for our process. Regions ABG and BGC are accessible when $x_1 > x_{2,3}$ and $x_2 > x_{1,3}$, respectively. Similarly, region AGC is accessible when $x_3 > x_{1,2}$.
In the dijet limit for thrust, where $\tau = 0$, only the line segment AB within region ABG becomes accessible for the first definition of thrust, T $= x_1$. This accessibility is limited to very small values of $\tau$. However, with increasing $\tau$, additional regions of ABG (away from AB) also become accessible. This trend extends to the second and third definitions of thrust, where areas other than line segment BC (AC) from region BGC (AGC) also become accessible as $\tau$ increases.

The key difference is, for the definition of thrust, the allowed phase space is around the region where $ x_i $ is largest, in the given definition of thrust. In contrast, for the case of spherocity precisely the opposite pattern is noticed, the allowed phase space region is around where $ x_i $ is smallest.
 Additionally, for thrust, however small $ \tau  $ becomes, the restricted phase regions do not overlap. Thus, the structure of contribution from region-I (ABG) and region-III (AGC) are different. However, for spherocity restricted phase space, we have noticed that these regions do overlap figures~\ref{fig:x1R1-R3@s=0.01} and \ref{fig:x1-x2_extraPLOTS}, as the value of $ s \to 0 $ the overlapping increases and at very small values ($ s =10^{-5} $) all the three curves entirely overlap which can be seen very clearly from figure~\ref{fig:x1-x2_extraPLOTS}\textcolor{blue}{a}. Due to this overlap, the structure of contribution from all three regions has a similar form, and we receive contributions at LLs and NLLs of LP and NLP from all three regions. However, these contributions from all three regions are not identical, as we have seen at higher values of $ s $, i.e. $ s=0.01 $ and $ 1/15 $, the curves shift towards edge where $ x_i $ is smallest, as previously discussed in black curve in figures~\ref{fig:x1R1-R3@s=0.01} and \ref{fig:x1-x2_extraPLOTS}\textcolor{blue}{b}. 

The contribution arising from region-III is more significant than regions-I and II contributions, and we notice that the fraction of LL at LP captured from region-III has more significant coefficients than the LL at LP from regions-I and II.
On the same note, in figure~\ref{fig:x3R1-R3}, we plot the phase space for $ x_3 $ against $ x_2 $ at three different values of $ s= 10^{-2} $ in three different regions. We observe a similar pattern here again, where we notice that curves from all three definitions cover almost the same parts of phase space, suggesting again that the third definition of spherocity must capture fractions of all the leading terms, like the LL at LP and others.
Upon varying the value of $ s $, we notice the contribution from the third region shifts towards the bottom-right part of the plot, at the bottom-right edge $ x_3 \to 0 $ and both $ x_2, \, x_1 \to 1 $. {This simple phase space analysis reveals that all three definitions of spherocity indeed contribute towards LP and NLP, and in the extreme dijet limit, i.e., $ s = 10^{-5} $, all three definitions become indistinguishable, and they overlap.}

\subsection{Plots for spherocity distribution}

\begin{figure}[h]
	\centering
	\includegraphics[width = 0.65\linewidth]{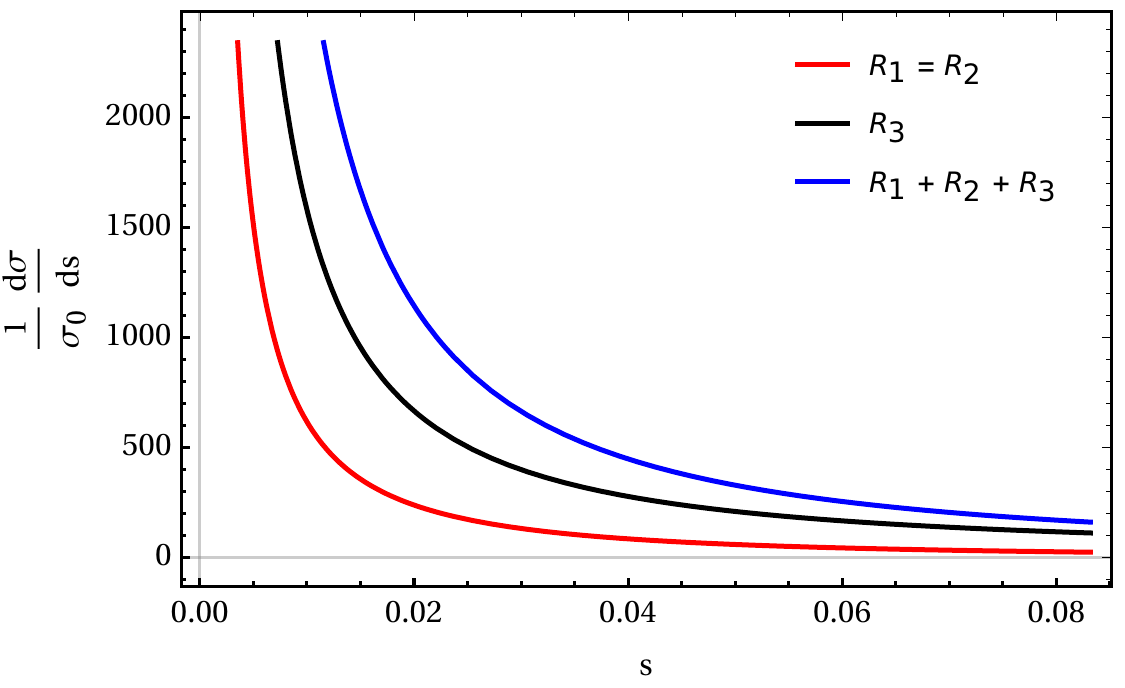}
	\caption{Plot for spherocity distribution against shape variable $ s $ as computed from three different regions, the red curve shows contribution from region-I which is identical to region-II's contribution. The black curve shows contribution from region-III while the blue curve shows the combined result from all the three regions. }
	\label{fig:NLOdstrR1-R3}
\end{figure}

This section explores the spherocity distribution by analyzing contributions from three distinct regions. The red curve in figure~\ref{fig:NLOdstrR1-R3} depicts the contribution solely from region-I, as defined by eq.~\eqref{eq:R1NLO}. Likewise, the black curve represents the contribution from region-III, as described in eq.~\eqref{eq:R3NLO}. Finally, the blue curve visualizes the combined contribution from all three regions, as given by eq.~\eqref{eq:NLO}.
We observe that the contribution from region-III is more significant than those from regions-I and II in the limit as $s \to 0$. This observation aligns with our findings from figures~\ref{fig:x1-x2_extraPLOTS}\textcolor{blue}{b} and \ref{fig:x3-x2_extraPLOTS}\textcolor{blue}{b}, where the third definition of spherocity is most effective at capturing the contributions arising from the limit $x_3 \to 0$. This is indicated by the difference in the red and black curves in figure~\ref{fig:NLOdstrR1-R3} for the spherocity distribution, with the black curve consistently at higher numerical values than the red curve throughout the allowed range of $s$. We have taken $\alpha_{s}= 0.1193$ at $\sqrt{s}=172$ GeV from LEP2, two loop result in~\cite{ALEPH:2003obs}, the numerical results are also given in~\cite{ALEPHsite}.\
Furthermore, the contribution from region-I is identical to that from region-II. Therefore, we only display the contribution from region-I in the red curve.

\section{Conclusion} \label{sec:conclusion}

Studying event shape variables in quantum chromodynamics has a long and rich history. It provides an organized framework for understanding the dynamics of strong interactions and probing the fundamental structure of matter. Through the analysis of event shape variables such as thrust, heavy jet mass, and jet broadening, the QCD predictions are made with high precision, obtaining valuable constraints on the strong coupling constant.
The next-to-leading power corrections in the next-to-leading order are required to increase the precision of theoretical predictions made by event shape studies, especially in reducing uncertainty in determining strong coupling constant. Additionally, including other event shape studies will further provide improved insights and solidify the predictions made by perturbative QCD. The primary reason for uncertainty in the determination of strong coupling constant from event shape prediction arises from hadronization, a non-perturbative phenomenon. As spherocity has found various applications in heavy-ion collisions as discussed previously, and its study can potentially serve as a bridge between the perturbative and non-perturbative QCD and help in constraining the uncertainties in the measurement of $ \alpha_{s} $ from event shape studies.
We compute the NLO distribution from three different regions based upon the definition of spherocity, where we were able to capture the LLs at at LP and NLP along with the NLLs from all three regions. The appearance of LL at LP from region-III is an exciting occurrence; the phase space constraint put forth by spherocity is different from thrust, and it explains the appearance of LLs at LP in this region. Thus, we believe studying the spherocity shape variable will help improve precision studies in perturbative QCD and further theoretical predictions made under the framework of QCD.

\section*{Acknowledgements}
The author would like to thank Anurag Tripathi for the essential discussions, Aditya Srivastav for reading the draft carefully and providing essential suggestions, and Swarna for her earlier presentation, in which the idea of this particular shape variable popped up. Additionally, the author acknowledges the support of CSIR, Govt. of India, for an SRF fellowship
(09/1001(0052)/2019-EMR-I).

\bibliographystyle{JHEP}
\bibliography{fullref}

\end{document}